\documentclass[twocolumn,showpacs,preprintnumbers,amsmath,nofootinbib,amssymb]{revtex4}

\usepackage{hyperref}
\usepackage{graphicx}
\usepackage{floatrow}
\usepackage{subfig}
\usepackage{dcolumn}
\usepackage{bm}

\def\be{\begin{equation}}
\def\ee{\end{equation}}
\def\bea{\begin{eqnarray}}
\def\eea{\end{eqnarray}}
\def\NO{\nonumber}
\def\gev{\mathrm{~GeV}}

\begin{document}


\title{The leptonic current structure and azimuthal asymmetry in deeply inelastic scattering}


\author{Hong-Fei Zhang$^{1,2}$}
\email{hfzhang@ihep.ac.cn (Corresponding Author)}

\author{Zhan Sun$^{3}$}
\email{zhansun@cqu.edu.cn (Corresponding Author)}

\affiliation{
$^{1}$ School of Science, Chongqing University of Posts and Telecommunications, Chongqing, 400065, China. \\
$^{2}$ Department of Physics, School of Biomedical Engineering, Third Military Medical University, Chongqing 400038, China. \\
$^{3}$ School of Science, Guizhou Minzu University, Guiyang, 500025, P. R. China.
}%
\date{\today}

\begin{abstract}
We present a compact form of the leptonic currents for the computation of the processes involving an initial virtual boson (photon, $W^{\pm}$ or $Z_0$).
For deeply inelastic scattering,
once the azimuthal angle of the plane expanded by the initial- and final-state leptons are integrated over in the boson-proton rest frame,
the azimuthal-asymmetric terms vanish, which, however,
is NOT true when some physical quantities (such as the transverse momentum of the observed particle) are specified in laboratory frame.
The abuse of the symmetry may lead to wrong results.
\end{abstract}

\pacs{12.38.Aw, 12.60.Hb, 13.85.Hd, 13.85.Ni}
\maketitle

\section{Introduction\label{sec:intro}}
\begin{figure}
\center{
\includegraphics*[scale=0.35]{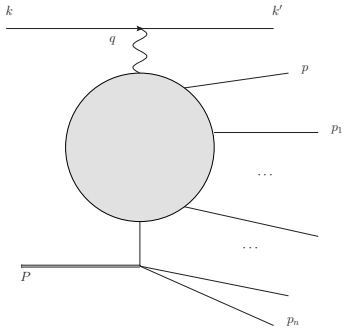}
\caption {\label{fig:diag}
The illustrative diagram for the DIS processes.
}}
\end{figure}

Deeply inelastic scattering (DIS) provides a useful probe to the partonic structure of the hadrons and photons.
It uses leptons to collide with the species of particles we study, and observes the kinematics of the scattered leptons,
which can provide information for the structure of the targets.
Taking the electron-proton scattering as an example,
the kinematics of the scattered leptons can be described by any two of the following four variables,
\bea
&&Q^2=-q^2\equiv-(k-k')^2,~~~~W^2=(P+q)^2, \NO \\
&&x=\frac{Q^2}{2P\cdot q},~~~~y=\frac{P\cdot q}{P\cdot k}, \label{eqn:lepvar}
\eea
where $P$, $k$, $k'$ and $q$ are, as illustrated in Figure~\ref{fig:diag}, the momenta of the initial proton,
the initial and final lepton, and the virtual boson, respectively.
The differential cross section can be written, in the limit $P^2/Q^2\ll 1$, as
\be
\mathrm{d}\sigma=\frac{1}{4P\cdot k}\frac{1}{N_cN_s}L_{\mu\nu}\frac{1}{(Q^2)^2}H^{\mu\nu}\mathrm{d}\Phi'\mathrm{d}\Phi_H, \label{eqn:cs}
\ee
where $1/(N_cN_s)$ is the color and spin average factor,
$L_{\mu\nu}$ and $H^{\mu\nu}$ are the leptonic and hadronic tensors, respectively, and
\bea
&&\mathrm{d}\Phi'=\frac{\mathrm{d}^3k'}{(2\pi)^32k'_0}, \NO \\
&&\mathrm{d}\Phi_h=\frac{\mathrm{d}^3p}{(2\pi)^32p_0}, \NO \\
&&\mathrm{d}\Phi_X=(2\pi)^4\delta^4(P+q-p-\sum_ip_i)\prod_i\frac{\mathrm{d}^3p_i}{(2\pi)^32p_{i0}}, \NO \\
&&\mathrm{d}\Phi_H=\mathrm{d}\Phi_h\mathrm{d}\Phi_X. \label{eqn:ps}
\eea
Here, $i$ runs over all the final states other than the scattered lepton and the tagged hadron ($p$).
The leptonic tensor can be directly calculated and obtained as
\bea
L_{\mu\nu}&=&4\pi\alpha\mathrm{Tr}(k\!\!\!\slash\gamma_\mu k'\!\!\!\!\slash\gamma^\nu) \NO \\
&=&8\pi\alpha Q^2[(-g_{\mu\nu}-\frac{q_\mu q_\nu}{Q^2})+\frac{(2k-q)_\mu(2k-q)_\nu}{Q^2}] \NO \\
&\equiv&8\pi\alpha Q^2l_{\mu\nu}, \label{eqn:lepcurrent}
\eea
where $\alpha$ is the electromagnetic coupling,
and the lepton mass and the contributions from the $Z_0$ propagator are neglected for the moment,
and will be considered in the sequel.
Having $d\Phi_H$ integrated over, the structure of the hadronic tensor,
\be
W^{\mu\nu}(P,q)\equiv\int H^{\mu\nu}(P,q,h,p_1,\ldots,p_n)\mathrm{d}\Phi_H, \label{eqn:wuv}
\ee
is restricted by the Lorentz covariance, thus,
can be decomposed into the linear combination of the current conserving dimensionless basic tensors as
\bea
W^{\mu\nu}&=&(-g^{\mu\nu}-\frac{q^\mu q^\nu}{Q^2})F_1(x,Q^2) \NO \\
&+&\frac{1}{Q^2}(q+\frac{Q^2}{P\cdot q}P)^\mu(q+\frac{Q^2}{P\cdot q}P)^\nu\frac{1}{2x}F_2(x,Q^2) \NO \\
&-&\frac{i}{P\cdot q}\epsilon^{\mu\nu\alpha\beta}P_\alpha q_\beta F_3(x,Q^2). \label{eqn:hadstruc}
\eea
For inclusive DIS at QCD leading order, $F_1$, $F_2$ and $F_3$ are dimensionless,
thus, independent of $Q^2$~\cite{Bjorken:1968dy, Bloom:1969kc, Breidenbach:1969kd}.
This scaling law indicates that, in high energy limit, the hadrons interact via pointlike partons inside them~\cite{Feynman:1969ej, Feynman:1973xc}.
The structure functions describe the parton distributions in high-energy hadrons (see e.g.~\cite{Buras:1979yt, Brock:1993sz}).

We thus can obtain
\bea
l_{\mu\nu}W^{\mu\nu}=2F_1(x,Q^2)+\frac{2-2y}{xy^2}F_2(x,Q^2). \label{eqn:contract}
\eea
It is easy to verify that, by setting
\bea
&&l_{\mu\nu}=\frac{2-2y+y^2}{y^2}\epsilon_{\mu\nu}-\frac{6-6y+y^2}{y^2}\epsilon_{L\mu\nu}, \label{eqn:lepreduced1}
\eea
or equivalently
\bea
&&l_{\mu\nu}=\frac{2-2y+y^2}{y^2}\epsilon_{T\mu\nu}-\frac{4(1-y)}{y^2}\epsilon_{L\mu\nu}, \label{eqn:lepreduced2}
\eea
where
\bea
&&\epsilon_{\mu\nu}=-g_{\mu\nu}-\frac{q_\mu q_\nu}{Q^2}, \NO \\
&&\epsilon_{L\mu\nu}=-\frac{1}{Q^2}(q_\mu+\frac{Q^2}{P\cdot q}P_\mu)(q_\nu+\frac{Q^2}{P\cdot q}P_\nu), \NO \\
&&\epsilon_{T\mu\nu}=\epsilon_{\mu\nu}-\epsilon_{L\mu\nu}, \label{eqn:eplt}
\eea
one can reproduce the contraction $L_{\mu\nu}H^{\mu\nu}$ for the real form of the leptonic tensor presented in Eq.(\ref{eqn:lepcurrent}).
Note that the leptonic momenta, $k$ and $k'$,
do not appear in the reduced leptonic tensor presented in Eq.(\ref{eqn:lepreduced1}) or Eq.(\ref{eqn:lepreduced2}),
the employment of which, thus, can greatly simplify the calculation of the cross sections.

\begin{figure}
  \begin{center}
  \subfloat[]{\includegraphics*[scale=0.3]{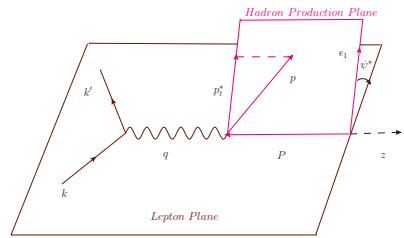} \label{fig:DISzframe}}\\
  \subfloat[]{\includegraphics*[scale=0.3]{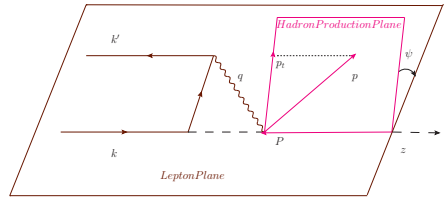} \label{fig:DISlab}}
  \end{center}
  \caption{
\label{fig:frames}
Illustrations of the process in $z$-frames (upper) and the laboratory frame (lower).
}
\end{figure}

For semi-inclusive DIS (SIDIS), an additional final state other than the scattered lepton is measured,
which will bring in additional scales.
Accordingly, the $Q^2$ scaling of the structure functions need to be counted.
It can be well described by the Dokshitzer-Gribov-Lipatov-Altarelli-Parisi Equations~\cite{Dokshitzer:1977sg, Gribov:1972ri, Altarelli:1977zs}.
In addition to the $Q^2$ scaling, transverse-spin and azimuthal asymmetry will also emerge in SIDIS
\cite{Georgi:1977tv, Cahn:1978se, Mendez:1978zx, Konig:1982uk, Sivers:1989cc, Cahn:1989yf, Sivers:1990fh, Chay:1991nh, Qiu:1991pp, Qiu:1991wg, Collins:1992kk, Collins:1993kq, Boer:1997nt, Qiu:1998ia},
which, in different kinematic regions, can provide crucial information for the transverse momentum dependent parton distribution
and the higher twist transverse-spin-dependent multiparton correlation functions.
On the experiment side, HERMES~\cite{Airapetian:2012yg} and COMPASS~\cite{Adolph:2014pwc} Collaborations measured the azimuthal asymmetries in SIDIS off unpolarized targets,
and observed nonvanishing cosine modulations and their strong dependence on the kinematical variables.
All the discussions in the literatures on the azimuthal angle and transverse spin are carried out in such frames in which the vector boson and the target travel along the opposite direction.
Without loss of generality, we assume the spacial momentum of the vector boson is along the $z$ direction,
and name such frames as $z$-frames (see Figure~\ref{fig:DISzframe}).
In any frame other than $z$-frames,
the azimuthal angle ($\psi$) dependence of the plane expanded by the initial and final leptons around the $z$-axis is not a simple trigonometric function,
and, with $\psi$ integrated over, does not vanish.
To our astonishment, numerous works employed the form of the leptonic tensor in Eq.(\ref{eqn:lepreduced1})
presenting results for the processes in which the transverse momentum ($p_t$) or the rapidity ($Y$) of the observed particle
in the laboratory frame is specified or a cut is applied on these parameters.
As a matter of fact, in the cases stated above, this form of the leptonic tensor will lead to wrong results.
Among these works\footnote{There are so many works did in the wrong framework, however, we only list one of them here.}
(see e.g.~\cite{Harris:1997zq}) are there some highly cited articles and calculations adopted by Monte Carlo generators.

In this paper, we will provide a compact form of the leptonic tensor,
which, on the one hand, is valid in all kinds of processes,
on the other hand, only involves momenta in hadronic process,
thus helps with the simplification of the computation.
In Section~\ref{sec:lepform}, we present the azimuthal dependent form of the leptonic tensor.
In Section~\ref{sec:comparison}, we discuss in detail the difference between the real form of the leptonic tensor and the reduced ones.
Section~\ref{sec:ee} is a short notes on the application of our approach in the $e^+e^-$ annihilation processes,
following which is the concluding remark in Section~\ref{sec:summary}.

\section{The Form of the Leptonic Tensor\label{sec:lepform}}

We will provide a compact form of the leptonic tensor for SIDIS,
and then a comprehensive form of the leptonic tensor for the most generalized DIS.
Having $\psi$ integrated over in $z$-frames, Eq.(\ref{eqn:lepreduced1}) will be automatically reproduced.
However, when some physical quantities, e.g. $p_t$ or $Y$ in the laboratory frame is specified,
the integration over $\psi$ (in any frame) does not lead to the conventional leptonic tensor in Eq.(\ref{eqn:lepreduced1}).
In the next section, we will see that the difference between the correct and the wrong results can be huge.

\subsection{The leptonic tensor for SIDIS}

Let us first investigate the processes in which only one final-state hadron ($p$) is observed.
Integrating over all the other hadronic final states, one can define the hadronic tensor
\bea
W_h^{\mu\nu}(P,q,p)\equiv\int H^{\mu\nu}(P,q,p,p_1,\ldots,p_n)\mathrm{d}\Phi_X. \label{eqn:whuv}
\eea
Since $W_h$ is a Lorentz covariant tensor and only dependent on the momenta, $P$, $q$, and $p$,
it can thus be decomposed as the linear combination of the tensors constituted of $-g^{\mu\nu}$, $P$, $q$, and $p$.
Note that $W_h$ satisfies the following equation
\bea
q_\mu W_h^{\mu\nu}=0. \label{eqn:qw}
\eea
Thus, with the above elements, only four independent tensors can be constructed.
One can define the normalized longitudinal and transverse vectors as
\bea
&&\epsilon_L=\frac{1}{Q}(q+\frac{Q^2}{P\cdot q}P), \NO \\
&&\epsilon_1=\frac{1}{p_t^\star}(p-\rho P-zq), \label{eqn:eplt}
\eea
where $p_t^\star$ is the transverse momentum of $p$ in $z$-frames,
$M$ is the mass of $p$,
$z=P\cdot p/P\cdot q$ is the elasticity coefficient, and
\be
\rho=\frac{p\cdot q+zQ^2}{P\cdot q}=\frac{p_t^{\star2}+M^2+z^2Q^2}{2zP\cdot q}. \label{eqn:rho}
\ee
We label all the physical quantities calculated in $z$-frames by superscript $\star$ hereinafter.
It is easy to check the following relations
\bea
&&q\cdot\epsilon_L=q\cdot\epsilon_1=P\cdot\epsilon_1=0, \NO \\
&&\epsilon_L^2=1,~~\epsilon_1^2=-1. \label{eqn:eplteqv}
\eea
As a matter of fact, $\epsilon_1$ is a normalized vector perpendicular to both $P$ and $q$,
and parallel to the transverse component of the momentum $p$ in $z$-frames,
as illustrated in Figure~\ref{fig:DISzframe}.
Accordingly, $W_h$ can be decomposed as
\bea
W_h^{\mu\nu}&=&W_g\epsilon^{\mu\nu}+W_L\epsilon_L^\mu\epsilon_L^\nu \NO \\
&+&W_{LT}(\epsilon_L^\mu\epsilon_1^\nu+\epsilon_1^\mu\epsilon_L^\nu)+W_T\epsilon_1^\mu\epsilon_1^\nu, \label{eqn:whSIDIS}
\eea
where $W_g$, $W_L$, $W_{LT}$ and $W_T$ are the coefficients of the four independent normalized tensors.

With a short calculation, one can obtain
\bea
l_{\mu\nu}W_h^{\mu\nu}&=&2W_g+\frac{4(1-y)}{y^2}W_L \NO \\
&-&\frac{4(2-y)}{y^2}\sqrt{1-y}\mathrm{cos}\psi^\star W_{LT} \NO \\
&+&[1+\frac{2-2y}{y^2}+\frac{2-2y}{y^2}\mathrm{cos}(2\psi^\star)]W_T, \label{eqn:lwh}
\eea
where $\psi^\star$ is the azimuthal angle of the lepton plane around the $z$-axis relative to the hadron production plane in $z$-frames,
as is illustrated in Figure~\ref{fig:DISzframe}.
It is easy to verify that, with
\bea
l^{\mu\nu}&=&A_g\epsilon^{\mu\nu}+A_L\epsilon_L^\mu\epsilon_L^\nu \NO \\
&+&A_{LT}(\epsilon_L^\mu\epsilon_1^\nu+\epsilon_1^\mu\epsilon_L^\nu)+A_T\epsilon_1^\mu\epsilon_1^\nu, \label{eqn:lepSIDIS}
\eea
where
\bea
&&A_g=1+\frac{2(1-y)}{y^2}-\frac{2(1-y)}{y^2}\mathrm{cos}(2\psi^\star), \NO \\
&&A_L=1+\frac{6(1-y)}{y^2}-\frac{2(1-y)}{y^2}\mathrm{cos}(2\psi^\star), \NO \\
&&A_{LT}=\frac{2(2-y)}{y^2}\sqrt{1-y}\mathrm{cos}\psi^\star, \NO \\
&&A_T=\frac{4(1-y)}{y^2}\mathrm{cos}(2\psi^\star), \label{eqn:coefSIDIS}
\eea
one can reproduce the results in Eq.(\ref{eqn:lwh}).
It seems as if the expressions of the reduced leptonic tensor in Eq.(\ref{eqn:lepSIDIS}) is more complicated than the real one in Eq.(\ref{eqn:lepcurrent}).
However, the reduced one only involves the momenta in the hadronic process;
in other words, the leptonic momenta do not appear.
Correspondingly, the number of the independent Lorentz invariants in the calculation,
while employing Eq.(\ref{eqn:lepSIDIS}), is reduced by 3,
which can greatly simplify the computation, especially when the calculation is carried out at loop level.

Taking into account the current conservation, the leptonic tensor in Eq.(\ref{eqn:lepSIDIS}) can be rewritten as
\bea
l^{\mu\nu}&=&C_1(-g^{\mu\nu})+C_2P^\mu P^\nu \NO \\
&+&C_3\frac{P^\mu p^\nu+p^\mu P^\nu}{2}+C_4p^\mu p^\nu, \label{eqn:lepSIDISc}
\eea
where
\bea
&&C_1=A_g, \NO \\
&&C_2=\frac{4x}{yS}(A_L-2\beta A_{LT}+\beta^2A_T), \NO \\
&&C_3=\frac{4x}{Qp_t^\star}(A_{LT}-\beta A_T), \NO \\
&&C_4=\frac{1}{p_t^{\star2}}A_T, \label{eqn:coefc}
\eea
with
\bea
\beta=\frac{p_t^{\star2}+M^2+z^2Q^2}{2zQp_t^\star}. \label{eqn:beta}
\eea
The leptonic tensor expressed in Eq.(\ref{eqn:lepSIDISc}) has replaced the normalized vectors by the momenta of the interacting particles,
which is more suitable for using in calculations.

The leptonic tensor is more complicated for charged and weak neutral current DIS,
when asymmetric tensors also participate.
This is true for the cases in which the beams and targets are polarized as well.
For most of the cases, the asymmetric part of the leptonic tensor is proportional to the following normalized structure,
\be
l^a_{\mu\nu}=\frac{2}{Q^2}\epsilon_{\mu\nu\alpha\beta}q^\alpha k^\beta, \label{eqn:lepasym}
\ee
and that of the hadronic tensor $W_h^{\mu\nu}$ can be decomposed as
\bea
W_h^{a\mu\nu}(P,q,p)=W^a_L\frac{1}{Q}\epsilon^{\mu\nu\alpha\beta}q_\alpha\epsilon_{L\beta} \NO \\
+W^a_T\frac{1}{Q}\epsilon^{\mu\nu\alpha\beta}q_\alpha\epsilon_{1\beta}. \label{eqn:wahSIDIS}
\eea
One can obtain
\bea
l^a_{\mu\nu}W_h^{a\mu\nu}=\frac{2(2-y)}{y}W^a_L-\frac{4}{y}\sqrt{1-y}\mathrm{cos}\psi^\star W^a_T. \label{lwah}
\eea
To reproduce the results in the above equation,
the leptonic tensor can be written in terms of the hadron momenta as
\bea
l^a_{\mu\nu}&=&\frac{2-y}{y}\frac{1}{Q}\epsilon_{\mu\nu\alpha\beta}q^\alpha\epsilon_L^\beta \NO \\
&+&\frac{2}{y}\sqrt{1-y}\mathrm{cos}\psi^\star\frac{1}{Q}\epsilon_{\mu\nu\alpha\beta}q^\alpha\epsilon_1^\beta. \label{eqn:lepasymSIDIS}
\eea

For inclusive DIS when the target is transversely polarized,
one can directly employ Eq.(\ref{eqn:lepSIDIS}) and Eq.(\ref{eqn:lepasymSIDIS}),
with $\epsilon_1$ replaced by the polarization vector of the target.

\subsection{The leptonic tensor for the most generalized DIS}

When several final states are observed and/or the polarization of the leptonic beam or the target is specified,
the hadron momenta and the polarization vectors are not constrained in a plane, accordingly,
the hadronic tensor is not only related to $P$, $q$, and $p$.
We need to introduce another vector, $\epsilon_2$, which,
in association with $\epsilon_L$ and $\epsilon_1$,
consists of a complete set of normalized orthogonal vectors perpendicular to $q$.
We define $\epsilon_2$ as
\bea
\epsilon_2^\mu=\frac{1}{Q}\epsilon^{\mu\nu\alpha\beta}q_\nu\epsilon_{L\alpha}\epsilon_{1\beta}, \label{eqn:ep2def}
\eea
where $\epsilon^{0123}=1$ in our convention.
Apparently, we have
\bea
q\cdot\epsilon_2=0=P\cdot\epsilon_2=p\cdot\epsilon_2=0,~~\epsilon_2^2=-1. \label{eqn:ep2eq}
\eea
Namely, $\epsilon_1$ and $\epsilon_2$ are the two normalized orthogonal transverse vectors in $z$-frames.
Any vector perpendicular to $q$ can be decomposed into the linear combination of $\epsilon_L$, $\epsilon_1$ and $\epsilon_2$.
Note that $2k-q$ is perpendicular to $q$.
To express $2k-q$ in terms of these three vectors, we need to calculate $2k\cdot\epsilon_i$ ($i=L, 1, 2$),
which are obtained as
\bea
&&2k\cdot\epsilon_L=Q(\frac{2}{y}-1), \NO \\
&&2k\cdot\epsilon_1=-\frac{2Q}{y}\sqrt{1-y}\mathrm{cos}\psi^\star, \NO \\
&&2k\cdot\epsilon_2=-\frac{2Q}{y}\sqrt{1-y}\mathrm{sin}\psi^\star. \label{eqn:kep}
\eea
Then, $2k-q$ can be expressed as
\bea
2k-q&=&Q(\frac{2}{y}-1)\epsilon_L+\frac{2Q}{y}\sqrt{1-y}\mathrm{cos}\psi^\star\epsilon_1 \NO \\
&+&\frac{2Q}{y}\sqrt{1-y}\mathrm{sin}\psi^\star\epsilon_2, \label{eqn:kdec}
\eea
employing which we can obtain the expression of $l^{\mu\nu}$ as
\bea
l^{\mu\nu}&=&A_1\epsilon_L^\mu\epsilon_L^\nu+A_2(\epsilon_L^\mu\epsilon_1^\nu+\epsilon_1^\mu\epsilon_L^\nu) \NO \\
&+&A_3(\epsilon_L^\mu\epsilon_2^\nu+\epsilon_2^\mu\epsilon_L^\nu)+A_4\epsilon_1^\mu\epsilon_1^\nu \NO \\
&+&A_5(\epsilon_1^\mu\epsilon_2^\nu+\epsilon_2^\mu\epsilon_1^\nu)+A_6\epsilon_2^\mu\epsilon_2^\nu, \label{eqn:eqlepgen}
\eea
where
\bea
&&A_1=\frac{4(1-y)}{y^2}, \NO \\
&&A_2=\frac{2(2-y)}{y^2}\sqrt{1-y}\mathrm{cos}\psi^\star, \NO \\
&&A_3=\frac{2(2-y)}{y^2}\sqrt{1-y}\mathrm{sin}\psi^\star, \NO \\
&&A_4=1+\frac{2(1-y)}{y^2}+\frac{2(1-y)}{y^2}\mathrm{cos}(2\psi^\star), \NO \\
&&A_5=\frac{2(1-y)}{y^2}\mathrm{sin}(2\psi^\star), \NO \\
&&A_6=1+\frac{2(1-y)}{y^2}-\frac{2(1-y)}{y^2}\mathrm{cos}(2\psi^\star), \label{eqn:coefgen}
\eea
and the relation,
\bea
-g^{\mu\nu}-\frac{q^\mu q^\nu}{Q^2}=-\epsilon_L^\mu\epsilon_L^\nu+\epsilon_1^\mu\epsilon_1^\nu+\epsilon_2^\mu\epsilon_2^\nu, \label{eqn:guv}
\eea
has been employed.
With $\psi^\star$ integrated from 0 to $2\pi$,
the form of the leptonic tensor presented in Eq.(\ref{eqn:lepreduced1}) or Eq.(\ref{eqn:lepreduced2}) can be reproduced.
However, note that the above statement is true only when the hadronic part of the cross section is independent of $\psi^\star$.

With Eq.(\ref{eqn:kdec}), the asymmetric tensor $l^a_{\mu\nu}$ can be obtained as
\bea
l^a_{\mu\nu}&=&\frac{2-y}{y}\frac{1}{Q}\epsilon_{\mu\nu\alpha\beta}q^\alpha\epsilon_L^\beta \NO \\
&+&\frac{2}{y}\sqrt{1-y}\mathrm{cos}\psi^\star\frac{1}{Q}\epsilon_{\mu\nu\alpha\beta}q^\alpha\epsilon_1^\beta \NO \\
&+&\frac{2}{y}\sqrt{1-y}\mathrm{sin}\psi^\star\frac{1}{Q}\epsilon_{\mu\nu\alpha\beta}q^\alpha\epsilon_2^\beta. \label{eqn:equivlepasym}
\eea
The integration over $\psi^\star$ leaves only the first term in Eq.(\ref{eqn:equivlepasym}),
which explicitly writes
\be
l^a_{\mu\nu}=\frac{2-y}{y}\frac{1}{Q}\epsilon_{\mu\nu\alpha\beta}q^\alpha\epsilon_L^\beta. \label{eqn:lepasymreduced}
\ee

Eq.(\ref{eqn:lepSIDIS}), Eq.(\ref{eqn:lepasymSIDIS}), Eq.(\ref{eqn:eqlepgen}) and Eq.(\ref{eqn:equivlepasym})
can reduce the number of the Lorentz invariants involved,
consequently, help to simplify the calculation.
Further, they can provide important information for the structure of the cross sections.

To complete our discussion, we note that, in collinear factorization,
one can simultaneously replace $P$ by the parton momentum in Eq.(\ref{eqn:eplt}) and Eq.(\ref{eqn:rho}).
Another issue to address is that Eq.(\ref{eqn:lepSIDIS}) can be reproduced by setting $A_3=A_5=0$ in Eq.(\ref{eqn:eqlepgen}),
while Eq.(\ref{eqn:lepasymSIDIS}) can be reproduced by dropping the last term on the right-hand side of Eq.(\ref{eqn:equivlepasym}).

\section{Comparison of the Reduced Leptonic Tensor to the Real One\label{sec:comparison}}

Eq.(\ref{eqn:lepreduced1}), Eq.(\ref{eqn:lepreduced2}) and Eq.(\ref{eqn:lepasymreduced})
are the generally used formulas in many literatures in the computation in DIS,
thus we name them as the conventional leptonic tensors.
However, these formulas are not valid when some quantities e.g. $p_t$ or $Y$ in the laboratory frame is specified.
This is generally because, as long as these quantities are specified,
the hadronic part of the cross section in NOT independent of $\psi^\star$,
which is actually manifest regarding the following relations,
\bea
&&p_t^2=p_t^{\star2}+z^2Q^2(1-y)-2zQp_t^\star\sqrt{1-y}\mathrm{cos}\psi^\star, \NO \\
&&Y=\frac{1}{2}\mathrm{ln}\{[p_t^{\star2}+M^2+z^2(1-y)Q^2 \NO \\
&&~~-2z\sqrt{1-y}Qp_t^\star\mathrm{cos}\psi^\star]/(4y^2z^2E_l^2)\}, \label{eqn:pty}
\eea
where $E_l$ is the energy of the incident lepton in the laboratory frame.
The derivation of Eq.(\ref{eqn:pty}) can be found in Appendix.
Once $p_t^2$ or $Y$ is specified,
the sine and cosine terms in Eq.(\ref{eqn:coefSIDIS}) and Eq.(\ref{eqn:coefgen}) do not vanish after the integration over $\psi^\star$.

To make our point clearer, we present here a more explicit form of the cross section for the case in which the $p_t$ in the laboratory frame is specified.
Here we constrain our discussions to only the symmetric leptonic tensor in SIDIS.
The differential cross section for one final-state hadron production can be expressed as (according to Eq.(\ref{eqn:cs}))
\bea
\mathrm{d}\sigma=\frac{1}{N_cN_s}\frac{4\pi\alpha}{SQ^2}l_{\mu\nu}W_h^{\mu\nu}\mathrm{d}\Phi'\mathrm{d}\Phi_h, \label{eqn:csSIDIS}
\eea
where
\bea
S=2P\cdot k, \label{eqn:s}
\eea
is the squared colliding energy.
The phase space can be obtained as
\bea
&&\mathrm{d}\Phi'=\frac{1}{32\pi^3}\mathrm{d}Q^2\mathrm{d}y\mathrm{d}\psi^\star, \NO \\
&&\mathrm{d}\Phi_h=\frac{1}{32\pi^3z}\mathrm{d}p_t^{\star2}\mathrm{d}z. \label{eqn:psred}
\eea
If we define
\bea
&&w_g=-g_{\mu\nu}W_h^{\mu\nu}, \NO \\
&&w_L=\epsilon_{L\mu}\epsilon_{L\nu}W_h^{\mu\nu}, \NO \\
&&w_{LT}=(\epsilon_{L\mu}\epsilon_{1\nu}+\epsilon_{1\mu}\epsilon_{L\nu})W_h^{\mu\nu}, \NO \\
&&w_T=\epsilon_{1\mu}\epsilon_{1\nu}W_h^{\mu\nu}, \label{eqn:wdef}
\eea
The differential cross section can then be expressed as
\bea
\mathrm{d}\sigma=\frac{\alpha}{256\pi^5N_sN_cSQ^2z}\sum_iA_iw_i\mathrm{d}Q^2\mathrm{d}y\mathrm{d}p_t^{\star2}\mathrm{d}z\mathrm{\psi^\star}, \label{eqn:csexp}
\eea
where $i$ runs over $g$, $L$, $LT$ and $T$.

Apparently, $A_i$ are functions of $y$ and $\psi^\star$,
while $w_i$ are functions of $Q^2$, $y$, $p_t^\star$, and $z$.
If one only measure the quantities in $z$-frames, e.g. $p_t^\star$ is fixed,
$w_i$ do not depend on $\psi^\star$, accordingly,
the integration over $\psi^\star$ makes the cosine terms in $A_i$ vanish.
However, if e.g $p_t$ in the laboratory frame is fixed,
the values of $p_t^\star$ and $\psi^\star$ are constrained in a curved surface.
When $\psi^\star$ varies, $w_i$ changes accordingly.
In this case, the cosine terms in $A_i$ will not vanish after the integration over $\psi^\star$.

To be more explicit, we can replace $\mathrm{d}p_t^{\star2}$ by $\mathrm{d}p_t^2$ with the Jacobian multiplied.
The Jacobian can be easily obtained regarding Eq.(\ref{eqn:pty}) as
\bea
J\equiv|\frac{\partial p_t^{\star2}}{\partial p_t^2}|=\frac{p_t^\star}{\sqrt{p_t^2-(1-y)z^2Q^2\mathrm{sin}^2\psi^\star}}. \label{eqn:jacobian}
\eea
Apparently, we have the following inequalities,
\bea
&&\int_0^{2\pi}d\psi^\star\mathrm{cos}\psi^\star J\neq0, \NO \\
&&\int_0^{2\pi}d\psi^\star\mathrm{cos}(2\psi^\star)J\neq0. \label{eqn:intcos}
\eea
Correspondingly, the conventional leptonic tensors cannot be reproduced with the integration over $\psi^\star$.

We can conclude that the structure functions, $F_1$ and $F_2$,
are not sufficient to describe the cross sections when $p_t$ or $Y$ are not taken to cover all their possible values.
Even for the processes in which $p_\mu P_\nu W_h^{\mu\nu}=p_\mu p_\nu W_h^{\mu\nu}=0$,
the conventional leptonic tensor in Eq.(\ref{eqn:lepreduced1}) or Eq.(\ref{eqn:lepreduced2}) will also lead to wrong results.

\begin{figure*}
\center{
\includegraphics*[scale=0.45]{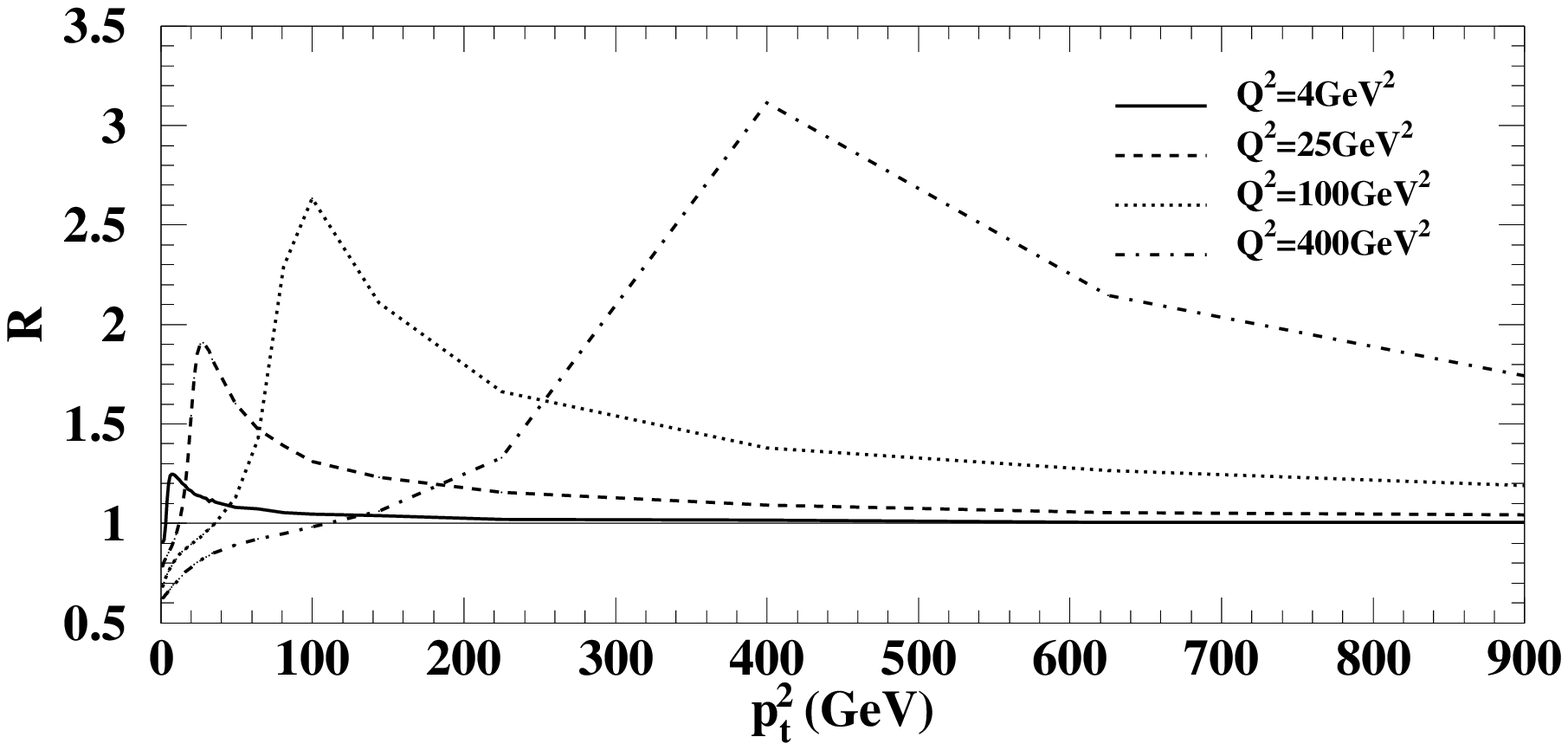}~~~~~~
\includegraphics*[scale=0.45]{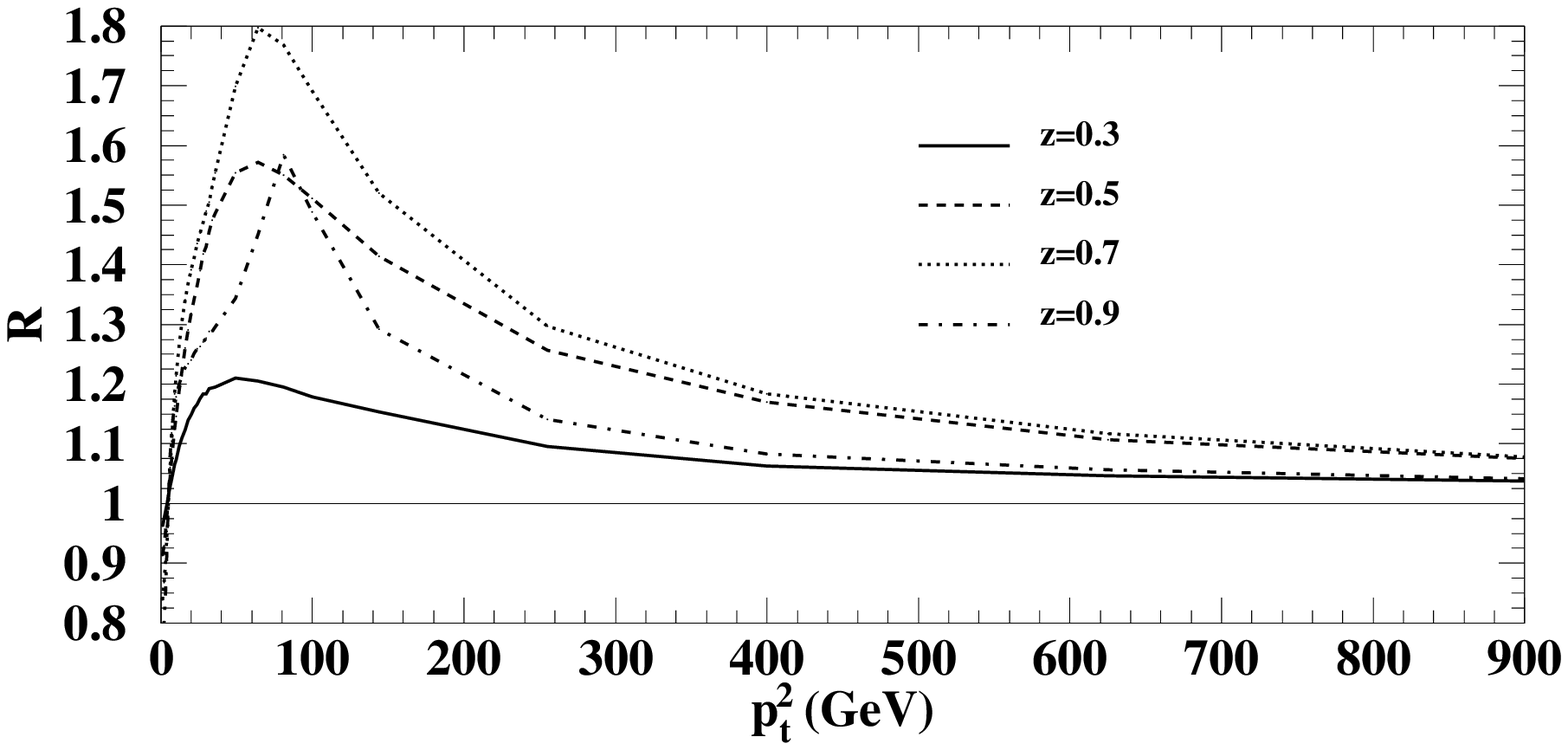}\\
\includegraphics*[scale=0.45]{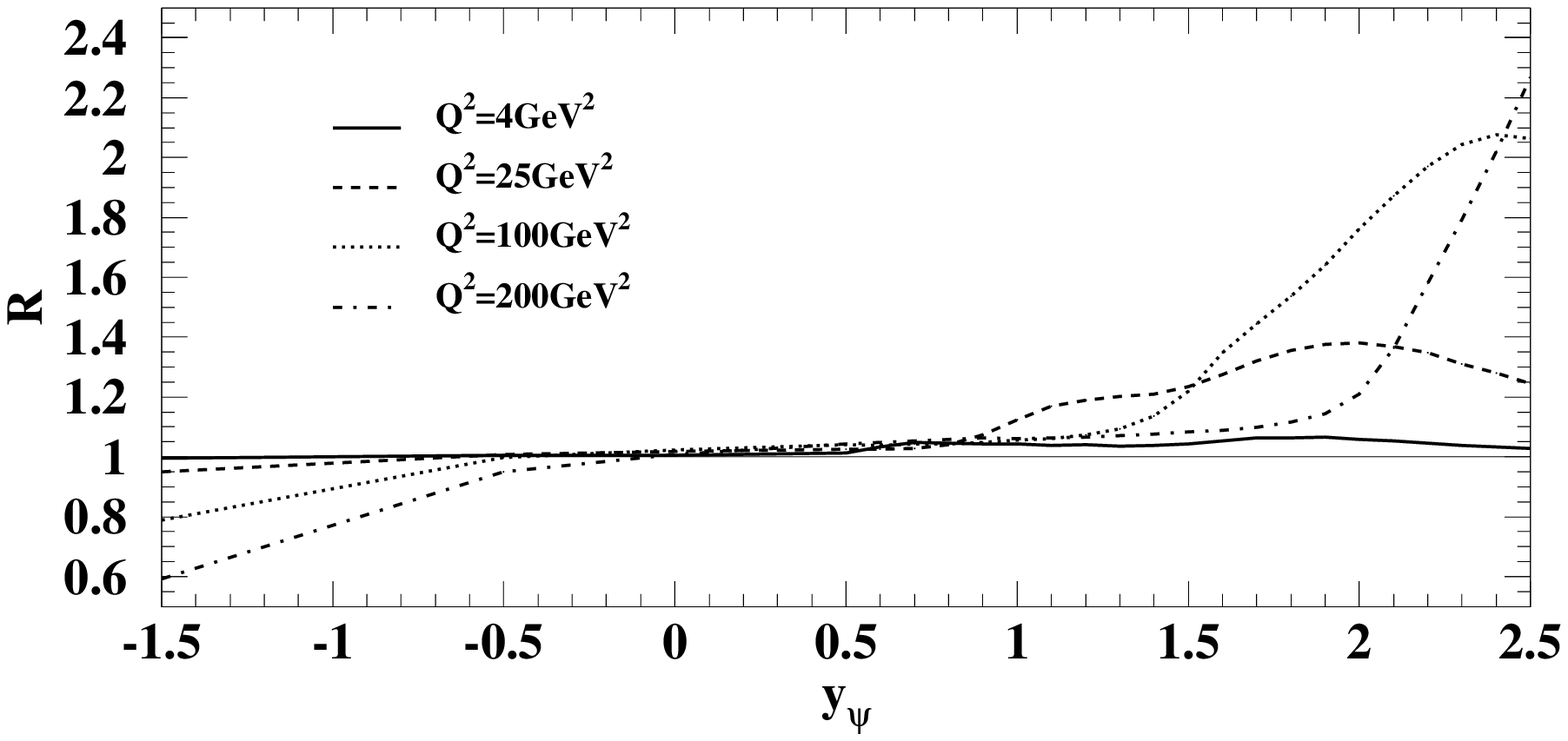}~~~~~~
\includegraphics*[scale=0.45]{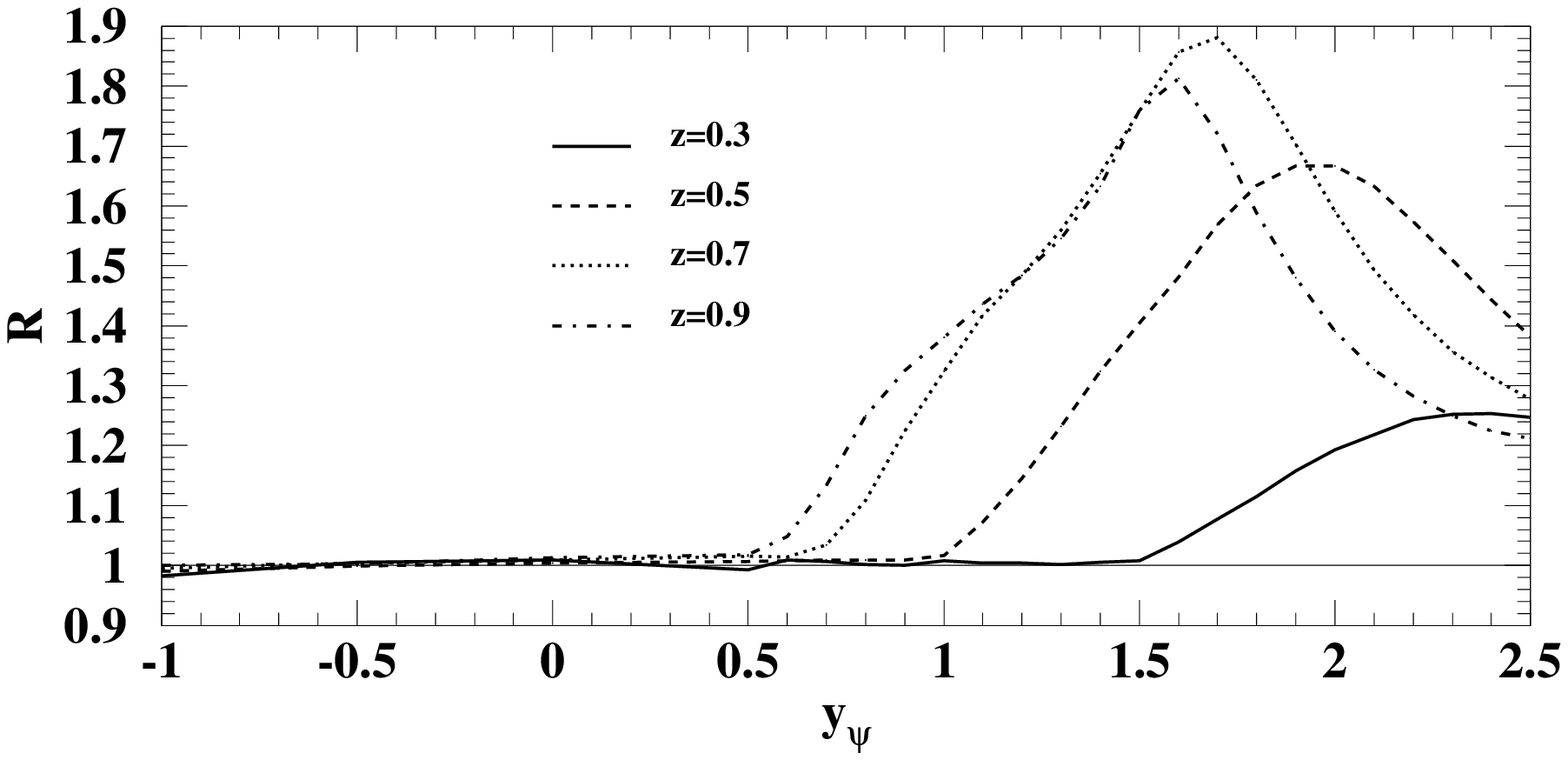}
\caption {\label{fig:ratio}
The ratio $R$ as a function of $p_t$ and $y_\psi$ for different values of $Q^2$ and $z$.
The invariant mass of the virtual photon and the proton lies in the region $60\gev<W<240\gev$.
For the left-hand-side plots, $0.3<z<0.9$, while for the right-hand-side plots, $4\gev^2<Q^2<100\gev^2$.
}}
\end{figure*}

We use $R$ to denote the ratio of the differential cross section calculated with the employment of the leptonic tensor
presented in Eq.(\ref{eqn:lepreduced1}), Eq.(\ref{eqn:lepreduced2}) and Eq.(\ref{eqn:lepasymreduced}) to the correct one,
say, that obtained using Eq.(\ref{eqn:lepcurrent}).
As an example, we investigate the $J/\psi$ inclusive production in photonic current DIS,
which has been studied in Ref.~\cite{Kniehl:2001tk}, however, in a wrong framework.
It is worth noting that with the same form of the leptonic tensor as given in Ref.~\cite{Kniehl:2001tk},
we can reproduce their results.
$R$ as functions of $p_t^2$ and the rapidity of the $J/\psi$ ($y_\psi$) in the laboratory frame are presented in Fig.\ref{fig:ratio}.
As $Q^2$ increases, the difference between the value of $R$ and 1 becomes larger in high $p_t$ region.
For $Q^2=400\gev^2$, the value of $R$ can be as large as 3.1 in high $p_t$ region and as small as 0.6 in low $p_t$ region.
For specified values of $z$ and integrated $Q^2$ in the range $4\gev^2<Q^2<100\gev^2$, the $p_t^2$ dependence of $R$ is also studied.
$R$ peaks at around $p_t^2\approx70\gev^2$ and $z\approx0.7$, where $R=1.8$ is obtained.
The difference between the wrong and correct results for specified values of $y_\psi$, however,
is not so large as that for specified values of $p_t^2$.
In midrapidity regions, especially $-0.5<y_\psi<0.5$, where most of the $J/\psi$ events are produced,
$R$ is almost equal to 1, which means in this region, even with the wrong form of the leptonic tensor,
one can generally reproduce the correct results.
However, this might be true only for specific processes.
For a different process, $R$ can be different from 1 even in midrapidity regions.

\section{Notes on the $e^+e^-$ Annihilation\label{sec:ee}}
The approach discussed above can also be applied to single inclusive particle production in $e^+e^-$ annihilation,
the squared amplitude (without averaging the initial spin) for which can be expressed as
\bea
|\mathcal{A}|^2=\frac{1}{s^2}L_{\mu\nu}H^{\mu\nu}. \label{eqn:ampsqee}
\eea
The hadronic tensor,
after the integration of all the phase space other than the observed particle and neglecting the contributions from the $Z$-boson propagator,
will be reduced into the combination of the following two tensors:
\bea
-g^{\mu\nu}+\frac{q^\mu q^\nu}{s},~~(p-\frac{p\cdot q}{s}q)^\mu(p-\frac{p\cdot q}{s}q)^\nu, \label{eqn:strucee}
\eea
where $q$ is the sum of the initial momenta, and $p$ is the momentum of the observed particle.
Taking the current conservation into account,
one can easily obtain
\bea
L_{\mu\nu}=4\pi\alpha s[-g_{\mu\nu}(1+\mathrm{cos}^2\theta)+\frac{p_\mu p_\nu}{\bm{p}^2}(1-3\mathrm{cos}^2\theta)], \label{eqn:lepee}
\eea
where $\theta$ is the angle between $\bm{p}$ and the spacial momentum of $e^-$ (or $e^+$) in the $e^+e^-$ center-of-mass frame.
Integrating over $\mathrm{cos}\theta$, the second term on the right-hand side of Eq.(\ref{eqn:lepee}) will vanish,
and the first term will reduce to $8\pi\alpha s/3$.
We can rewrite Eq.(\ref{eqn:lepee}) in the Lorentz invariant form by substituting the following equations,
\bea
\bm{p}^2=\frac{(p\cdot q)^2}{s}-M^2,~~\mathrm{cos}^2\theta=\frac{(p\cdot k-p\cdot k')^2}{\bm{p}^2s}, \label{eqn:pcos}
\eea
where $k$ and $k'$ are the momenta of $e^-$ and $e^+$, respectively, and $M$ is the mass of $p$.

\section{Summary\label{sec:summary}}

In summary, we investigated the structure of the leptonic tensors in DIS and SIDIS.
The most general forms of the leptonic tensor are presented in Eq.(\ref{eqn:eqlepgen}), Eq.(\ref{eqn:coefgen}), and Eq.(\ref{eqn:equivlepasym}),
which can greatly simplify the computation by reducing the number of the independent Lorentz invariants.
Moreover, they explicitly prove the azimuthal structure of the cross sections consists, in addition to the $\psi^\star$ independent terms,
of those proportional to $\mathrm{cos}\psi^\star$, $\mathrm{sin}\psi^\star$, $\mathrm{cos}(2\psi^\star)$, and $\mathrm{sin}(2\psi^\star)$, respectively,
which after the integration over $\psi^\star$ will vanish.
For SIDIS, the symmetric leptonic tensor reduces to a more compact form, which only involve four independent normalized tensors,
while the asymmetric one consists of only two independent normalized tensors.
Taking the $J/\psi$ inclusive production in DIS as an example,
we demonstrate that the reduced formalism of the leptonic tensor presented in Eq.(\ref{eqn:lepreduced1}) or Eq.(\ref{eqn:lepreduced2})
(as well as in Eq.(\ref{eqn:lepasymreduced})) cannot give correct predictions when some physical observables,
such as $p_t$ and the rapidity, in laboratory frame are measured.
However, many works are still using these reduced leptonic currents in computations.
Our work can provide a reference for the future phenomenological studies in DIS, including the physics at the future Electron-Ion Collider (EIC).

This work is supported by the National Natural Science Foundation of China (Grant No. 11405268 and 11647113).

\appendix

\section{The derivation of the relation between $p_t$ and $p_t^\star$}

Since $p_t$ is invariant under the boost along the $z$-axis,
we thus parameterize all the momenta in virtual-photon-proton ($\gamma^\star p$) rest frame and the laboratory frame.

The invariants $k\cdot p$ and $P\cdot p$ can be expressed in the laboratory frame as
\bea
&&k\cdot p=E_l\sqrt{p_t^2+M^2}e^Y, \NO \\
&&P\cdot p=E_p\sqrt{p_t^2+M^2}e^{-Y}, \label{eqn:kppp}
\eea
where $E_l$ and $E_p$ are the momenta of the incident lepton and proton, respectively,
and the forward $z$ direction is defined as that of the incident proton.
Then we have
\bea
4(k\cdot p)(P\cdot p)=4E_lE_p(p_t^2+M^2)=S(p_t^2+M^2). \label{eqn:mt2s}
\eea
$2P\cdot p$ can also be obtained in terms of the hadronic variables as
\bea
2P\cdot p=yzS, \label{eqn:pdp}
\eea
thus, we have
\bea
p_t^2+M^2=2yzk\cdot p. \label{eqn:mt2}
\eea

Now, we calculate $2k\cdot p$.
Apparently, it depends on $\psi^\star$.
To obtain its explicit expression, we need to parameterize the momenta in the $\gamma^\star p$ rest frame.
In the $\gamma^\star p$ rest frame, the forward $z$ direction is defined as that of the incident virtual photon,
which is consistent with the HERA experiment conventions.
Since the invariant energy of the $\gamma^\star p$ system is $W$,
we can obtain their energies and longitudinal momenta as
\bea
&&P_0^\star=\frac{W^2+Q^2}{2W}, \NO \\
&&P_l^\star=-\frac{W^2+Q^2}{2W}, \NO \\
&&q_0^\star=\frac{W^2-Q^2}{2W}, \NO \\
&&q_l^\star=\frac{W^2+Q^2}{2W}. \label{eqn:Pqs}
\eea
If we define $k^\mu$ as
\bea
k^\mu=(k_0^\star, \bm{k_t}^\star, k_l^\star),
\eea
we can calculate $2k\cdot P$ and $2k\cdot q$ as
\bea
&&2k\cdot P=S=\frac{W^2+Q^2}{2W}(k_0^\star+k_l^\star), \NO \\
&&2k\cdot q=-Q^2=\frac{W^2+Q^2}{2W}k_l^\star-\frac{W^2-Q^2}{2W}k_0^\star. \label{eqn:kpkq}
\eea
Then we can obtain $k_0^\star$ and $k_l^\star$ as
\bea
&&k_0^\star=\frac{S-Q^2}{2W}, \NO \\
&&k_l^\star=\frac{1}{2W}(Q^2+\frac{W^2-Q^2}{W^2+Q^2}S). \label{eqn:k0kls}
\eea
$k_t^{\star2}$ can be calculated via
\bea
k_t^{\star2}=k_0^{\star2}-k_l^{\star2}=Q^2\frac{1-y}{y^2}. \label{eqn:kts}
\eea
$p$ can be expressed in the $\gamma^\star p$ rest frame as
\bea
p^\mu=(m_t^\star(e^{Y^\star}+e^{-Y^\star}), \bm{p_t}^\star, m_t^\star(e^{Y^\star}-e^{-Y^\star})), \label{eqn:pmu}
\eea
where
\bea
m_t^\star=\sqrt{p_t^{\star2}+M^2}. \label{eqn:mts}
\eea
Then we can obtain $2k\cdot p$ as
\bea
2k\cdot p&=&\frac{1}{yz}[p_t^{\star2}+M^2+(1-y)z^2Q^2 \NO \\
&-&2z\sqrt{1-y}Qp_t^\star\mathrm{cos}\psi^\star]. \label{eqn:2kp}
\eea
With Eq(\ref{eqn:mt2}), we obtain
\bea
p_t^2=p_t^{\star2}+(1-y)z^2Q^2-2z\sqrt{1-y}Qp_t^\star\mathrm{cos}\psi^\star.
\eea


\end{document}